\documentclass[preprint]{aastex}  %% For preprint style
\shortauthors{Fassnacht et al.}
\shorttitle{Time delays in CLASS B1608+656}

\newcommand{\rahr}{$^{\rm h}$}
\newcommand{\ramn}{$^{\rm m}$}
\newcommand{\ksm}{km\ s$^{-1}$\,Mpc$^{-1}$}

\newcommand{\plotfour}[4]{
{\plottwo{{#1}}{{#2}}}

{\plottwo{{#3}}{{#4}}}
}

\begin{document}

\title{A Determination of $H_0$ with the CLASS Gravitational Lens B1608+656:
III.~A Significant Improvement in the Precision of the Time Delay Measurements}

\author{C. D. Fassnacht}
\affil{
   Space Telescope Science Institute,
   3700 San Martin Drive, 
   Baltimore, MD 21218; \\
      and
   National Radio Astronomy Observatory,
   P.O.\ Box O,
   Socorro, NM 87801
}
\email{cdf@stsci.edu}

\author{E. Xanthopoulos}
\affil
{
   University of Manchester, 
   Jodrell Bank Observatory,
   Macclesfield, 
   Cheshire SK11 9DL, UK.
}
\email{emily@jb.man.ac.uk}

\author{L. V. E. Koopmans}
\affil
{
   Theoretical Astrophysics, 
   California Institute of Technology,
   130-33,
   Pasadena, CA 91125.
}
\email{leon@tapir.caltech.edu}

\and

\author{D. Rusin}
\affil
{
   Harvard-Smithsonian Center for Astrophysics,
   60 Garden Street,
   Cambridge, MA  02138.
}
\email{drusin@cfa.harvard.edu}

\begin{abstract}

The gravitational lens CLASS B1608+656 is the only four-image lens
system for which all three independent time delays have been measured.
This makes the system an excellent candidate for a high-quality
determination of $H_0$ at cosmological distances.  However, the
original measurements of the time delays had large (12--20\%)
uncertainties, due to the low level of variability of the background
source during the monitoring campaign.  In this paper, we present
results from two additional VLA monitoring campaigns.  In contrast to
the $\sim$5\% variations seen during the first season of monitoring,
the source flux density changed by 25--30\% in each of the subsequent
two seasons.  We analyzed the combined data set from all three seasons
of monitoring to improve significantly the precision of the time delay
measurements; the delays are consistent with those found in the
original measurements, but the uncertainties have decreased by factors
of two to three.  We combined the delays with revised isothermal mass
models to derive a measurement of $H_0$.  Depending on the positions
of the galaxy centroids, which vary by up to 0\farcs1 in HST images
obtained with different filters, we obtain $H_0 =$ 61--65 \ksm, for
$(\Omega_M,\Omega_\Lambda) = (0.3,0.7)$.  The value of $H_0$ decreases
by 6\% if $(\Omega_M,\Omega_\Lambda) = (1.0,0.0)$.  The formal
uncertainties on $H_0$ due to the time delay measurements are $\pm 1$
($\pm 2$) \ksm\ for the 1$\sigma$ (2$\sigma$) confidence limits.
Thus, the systematic uncertainties due to the lens model, which are on
the order of $\pm$15 \ksm, now dominate the error budget for this
system.  In order to improve the measurement of $H_0$ with this lens,
new models that incorporate the constraints provided by stellar
dynamics and the optical/infrared Einstein ring seen in HST images
must be developed.

\end{abstract}


\section{Introduction}

Gravitational lenses provide excellent tools for the study of
cosmology.  In particular, the method developed by \citet{refsdal} can
be used to determine the Hubble Constant at cosmological distances.
This method requires a lens system for which both lens and source
redshifts have been measured, for which a well-constrained model of
the gravitational potential of the lensing mass distribution has been
determined, and for which the time delays between the lensed images
have been measured.  To date, measurements of time delays have been
reported in the literature for 11 lens systems: 
0957+561        \citep{tk0957_1,tk0957_2}, 
PG~1115+080     \citep{pls1115,barkana1115}, 
JVAS~B0218+357  \citep{adb0218,cohen0218}, 
PKS~1830-211    \citep{lovell1830,tw1830},
CLASS~B1608+656 \citep[][hereafter Paper I]{paper1}, 
CLASS~B1600+434 \citep{lvek1600,burud1600}, 
JVAS~B1422+231  \citep{1422delay},
HE~1104-1805    \citep[see, however, Pelt, Refsdal, \& Stabell 
 2002]{gilmerino1104},
HE~2149-2745    \citep{burud2149},
RX~J0911.4+0551 \citep{hjorth0911}, and
SBS~1520+530    \citep{burud1520}.  
Of these lens systems, CLASS B1608+656 is the only
four-image system for which all three independent time delays have
been unambiguously measured.  

\begin{figure}
\plotone{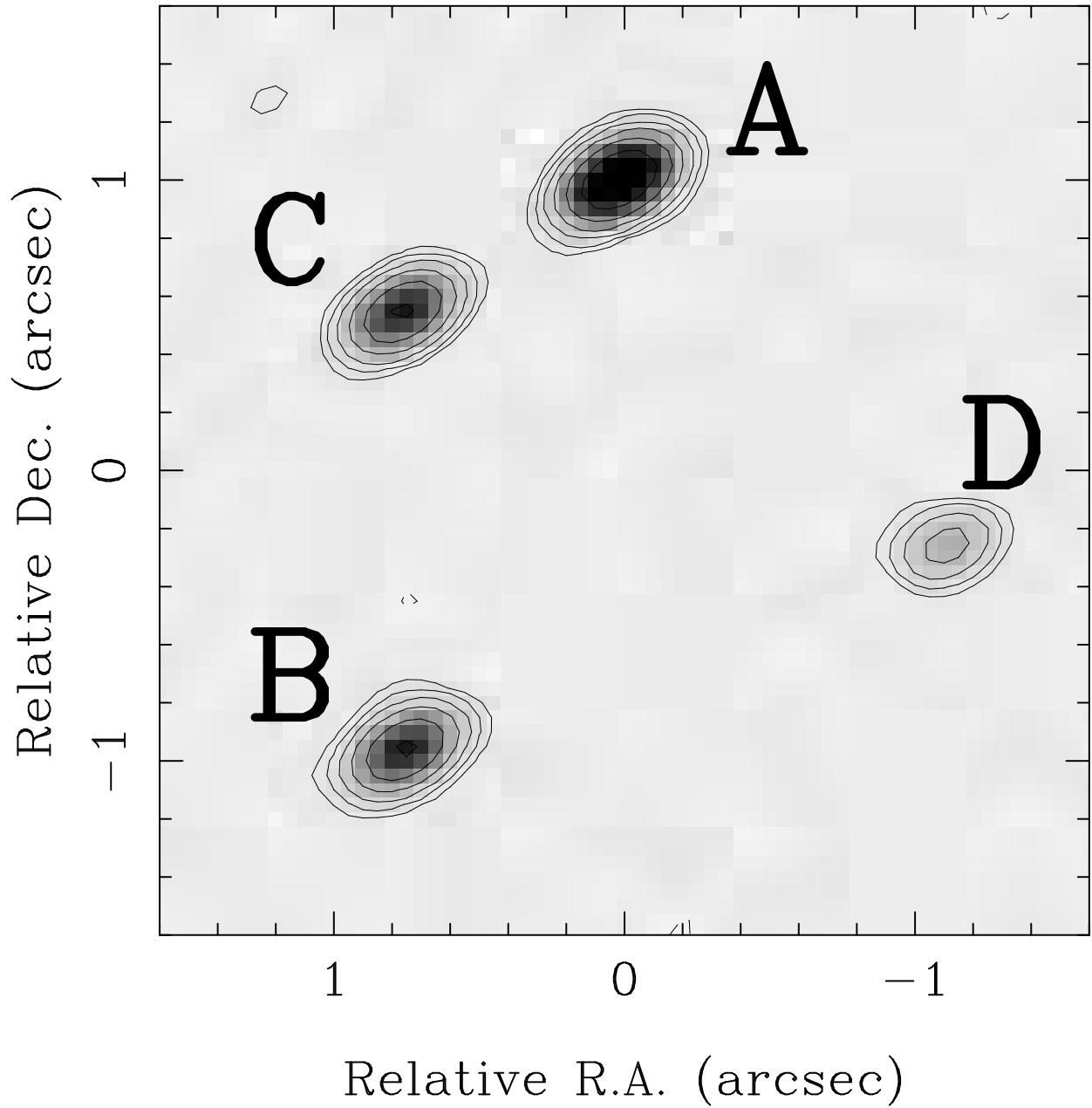}
\caption{Typical radio map of B1608+656 from the season 3 monitoring
campaign.  These data were obtained on 1999 August 08.  Both the
grayscale and the contours represent radio brightness.  The contours
are set at $-$2.5, 2.5, 5, 10, 20, 40, 80, and 160 times the RMS
noise level of 0.13 mJy~beam$^{-1}$.
\label{fig_1608map}}
\end{figure}

The CLASS B1608+656 lens system consists of the core of a radio-loud
poststarburst galaxy at a redshift of $z = 1.394$ \citep{zs1608} being
lensed by a $z = 0.630$ pair of galaxies \citep{sm1608}.  Radio maps
of the system show four images of the background source arranged in a
typical lens geometry (Figure~\ref{fig_1608map}).
Models of the lens system predict that, if the background source is
variable, image B should show the variation first, followed by
components A, C, and D in turn \citep[hereafter
Paper~II]{sm1608,paper2}.  The time delays determined in Paper I were
based on radio-wavelength light curves obtained with the Very
Large Array
(VLA\footnote{The National Radio Astronomy Observatory is a facility of
the National Science Foundation operated under cooperative agreement
by Associated Universities, Inc.})  between 1996 October and 1997 May.
During that time, the background source 8.5~GHz flux density varied by
$\sim$5\%.  Although the measured time delays were robust, the small
level of variability meant that the uncertainties on the delays were
large.  The estimated uncertainties from the Paper I data ranged from
$\sim$12\% for the B$\rightarrow$D time delay ($\tau_{BD}$) to
$\sim$20\% for the other two measured delays (95\% CL).  These
uncertainties translated to uncertainties on the derived value of the
Hubble Constant of $\sim$15\% (Paper~I).  In addition, the lens
modeling contributes an uncertainty of $\sim$30\% to the derived
$H_0$, with the largest contribution coming from the uncertainty in
the radial profile of the mass distribution in the lensing galaxies
(Paper~II).  In order to achieve a more precise determination of $H_0$
from this lens system, the uncertainties on both the time delays and
the modeling need to be significantly reduced.  This paper addresses
the first of those issues.

\section{Observations}

In VLA test observations conducted before the start of the first
monitoring program, the flux density of the B1608+656 components had
varied by up to 20\% on time scales of months.  If detected during a
monitoring program, variations on this scale would allow a significant
reduction in the time delay uncertainties.  Thus, we continued
our monitoring program with the VLA.  Our observations were conducted
at 8.5~GHz.  The angular resolution required to separate properly the
individual images of the background source limited the observations to
the times in which the VLA was in its most extended configurations,
the A, BnA, and B arrays.  Thus, we have light curves covering the
approximately eight months out of every 16-month cycle, during which
the VLA was in the proper configurations.  We have denoted these
eight-month blocks as observing ``seasons,'' with season 1 being 
1996 November to 1997 May, which was discussed in Paper I.

In this paper, we present results from seasons 2 and 3.
Table~\ref{tab_obs} summarizes the observations that were conducted.
Each observing epoch contained observations of CLASS B1608+656; the
phase calibrator, 1642+689; the primary flux calibrator, 3C\,343
(1634+628); and secondary flux calibrators.  The phase center for the
observations of B1608+656 was chosen to be the approximate center of
the system at RA: 16\rahr09\ramn13\fs953 and Dec:
+65\degr32\arcmin28\farcs00 (J2000).  Our primary flux calibrator is
not a standard VLA flux calibrator.  However, it is much closer in the
sky to our lens system than the usual flux density calibrators,
3C\,286 and 3C\,48, and our season 1 observations showed that its flux
density is stable (Paper I).  Any errors in the absolute flux-density
calibration due to either variability in this source or secondary
instrumental effects could be corrected through the use of the
secondary flux calibrators.  These secondary calibrators are
steep-spectrum radio sources and, as such, should not be highly
variable.  In particular, if the light curves of the secondary flux
calibrators show variations that are correlated, it is almost certain
that these variations are produced by flux calibration errors rather
than being intrinsic.  Thus, the information obtained from correlated
variability of the secondary calibrators can be used to correct the
light curves of the lensed images (see Paper~I; also
\S\ref{sec_finalcurves}).  Compact symmetric objects (CSOs) have been
shown to be especially stable secondary calibrators at 8.5~GHz
\citep[hereafter FT01]{csocals}.  The season 2 secondary flux
calibrators were 1633+741, which was used in the Paper I analysis, and
two CSOs: J1400+6210 and J1944+5448.  The integration times on the
various sources were typically one minute on the phase calibrator and
the CSOs, two minutes on 1633+741, and approximately five minutes on
the lens system.  Neither CSO was observed for the full length of
season 2; J1944+5448 was observed from 1998 February 13 to 1998 June
01 while J1400+6210 was observed from 1998 May 12 to 1998 October 19.
The observations for season 3 were similar to those of season 2, but
with a different set of secondary flux calibrators.  We dropped
1633+741 and J1944+5448 and added six CSOs to the two used in the
season 2 observations.  The increase in the number of good secondary
flux calibrators provided a better estimate of the errors in the
absolute flux density calibration.  The additional CSOs used were:
J1035+5628, J1148+5924, J1244+4048, J1545+4751, J1823+7938, and
J1945+7055.  Because there were more sources observed per epoch than
in the first two seasons, the average integration time per source
dropped slightly.  The phase calibrator and CSOs typically were
observed for 45-60~sec and B1608+656 was observed for a few minutes.

\begin{deluxetable}{cllccc}
\tablenum{1}
\tablewidth{0pt}
\scriptsize
\tablecaption{Monitoring Observations\label{tab_obs}}
\tablehead{
\colhead{Season}
 & \colhead{Start Date}
 & \colhead{End Date}
 & \colhead{$N_{\rm obs}$}
 & \colhead{$N_{\rm good}$}
 & \colhead{Average Spacing (d)}
}
\startdata
1 & 1996 Oct 10 & 1997 May 26 & 66 & 64 & 3.6 \\
2 & 1998 Feb 13 & 1998 Oct 19 & 81 & 78 & 3.1 \\
3 & 1999 Jun 15 & 2000 Feb 14 & 92 & 88 & 2.7 \\
\enddata
\end{deluxetable}

\section{Data Reduction}

\subsection{Calibration}

The data were calibrated using customized procedures that acted as an
interface to tasks in the NRAO data reduction package, AIPS.  For
consistency, the data from all three observing seasons were calibrated
in the same way.  Therefore, the season 1 data were recalibrated.  The
overall flux density calibration was tied to 3C\,343.  As discussed in
FT01, 3C\,343 is not an ideal calibrator due to some low
surface-brightness emission surrounding the central component.
However, it was possible to correct for the changes in the total
measured flux density of 3C\,343 as the VLA configuration changed (see
\S\ref{sec_finalcurves}).

The calibration procedure for each epoch of observation began with
careful data editing.  Bad integrations on the calibrators, indicated
by discrepant amplitudes or phases, were deleted.  Any antenna which
exhibited a large number of bad integrations during an epoch was
flagged for the entire epoch.  The antenna system temperatures were
also used as a diagnostic to discover bad integrations, which were
then flagged.  One antenna in particular (antenna 3) had system
temperatures that were, on average, a factor of three higher than
those of the other antennas.  Because this behavior was seen during
all three seasons of observations, we flagged antenna 3 before doing
any calibration.  After the bad data had been flagged, the antenna
gain phases and amplitudes were determined through self-calibration on
the bright, compact calibrator sources.  For all of the CSOs and the
phase calibrator, we used a point source model to do the self
calibration.  The emission from two of the CSOs, J1035+5628 and
J1400+6210, is slightly resolved so, for those sources, we imposed a
maximum baseline length of 400~k$\lambda$ in the calculation of the
gain solutions.  The more complicated morphology of the flux
calibrator required a simple model as the input to the
self-calibration.  Otherwise, the processing for the flux calibrator
was the same as that used for the phase calibrator and CSOs.  The
self-calibration was conducted in two steps.  In the first step,
solutions were calculated for the gain phases only, with a solution
interval of 10~sec.  Those solutions were applied to the data and then
the second calibration was performed.  This time, both phase and
amplitude solutions were calculated, with a solution interval of
30~sec.  The flux-density scale was then transferred to the CSOs and
phase calibrators through the GETJY task.  Finally, the gain solutions
from the phase calibrators were interpolated and applied to the data
on B1608+656 and, for seasons 1 and 2, 1633+741.

\subsection{Mapping and Initial Light Curves}

The flux densities for the B1608+656 components, the CSOs, and
1633+741 were obtained through the use of the model-fitting procedures
in the {\tt DIFMAP} software package \citep{difmap}.  The process used
to derive the flux densities for the CSOs was described in
FT01.  The flux densities for B1608+656 and 1633+741 were
measured using methods similar to those described in Paper I, but with
small variations.  For 1633+741 the only change in the processing was
to limit the $(u,v)$ range used for the model-fitting procedure to
include only baselines with lengths less than 400~k$\lambda$.  This
change de-emphasized the contribution of small-scale structure in the
source, namely the emission associated with the probable core
component.  The data on the source suggested that the core of 1633+741
was variable during the observations.  By minimizing the contribution
of the core in the model-fitting, we obtained light curves which had
less intrinsic scatter, especially in the A-configuration data, and
which were much more consistent with the CSO light curves.  

For B1608+656, we investigated the effect of letting the component
positions vary during the model fitting.  We ran the DIFMAP scripts
twice, once with variable positions and once with the positions fixed
at the values given in Table~\ref{tab_compinfo}.  For the case in
which we let the positions vary, the best-fit positions returned by
the model-fitting were very close to those in the fixed-position
models.  For all components the mean offsets from the fixed
positions were 1 milliarcsecond (mas) or less.  The RMS scatter about the
mean ranged from 1 to 4~mas, with the largest RMS offsets being those
associated with component D in seasons 2 and 3, when the component was
faintest.  The offsets are in all cases small compared to the
VLA beam sizes of 250 to 700~mas.
As expected for such small positional differences, there were no
significant differences in the flux densities obtained via the two
fitting procedures.  Thus, we chose to use only the results obtained
by holding the positions fixed.

\begin{deluxetable}{crrrrr}
\tablenum{2}
\tablewidth{0pt}
\scriptsize
\tablecaption{Component Positions and Mean Flux Densities\label{tab_compinfo}}
\tablehead{

 & \colhead{$x$\tablenotemark{a}}
 & \colhead{$y$\tablenotemark{a}}
 & \colhead{$<S_1>$}
 & \colhead{$<S_2>$}
 & \colhead{$<S_3>$} \\
\colhead{Component}
 & \colhead{(arcsec)}
 & \colhead{(arcsec)}
 & \colhead{(mJy)}
 & \colhead{(mJy)}
 & \colhead{(mJy)}
}
\startdata
A &    0.000 &    0.000 &    34.1 &    21.8 &    22.8 \\
B & $-$0.738 & $-$1.962 &    16.8 &    10.6 &    11.6 \\
C & $-$0.745 & $-$0.453 &    17.3 &    11.3 &    11.6 \\
D &    1.130 & $-$1.257 & \phn5.9 & \phn4.0 & \phn3.7 \\
\enddata
\tablenotetext{a}{Component positions are given in Cartesian rather than
astronomical coordinates.}
\end{deluxetable}

\subsection{Light Curve Editing}

During the course of the observations, there were some data sets
obtained under less-than-ideal conditions.  Although VLA observations
at 8.5~GHz are relatively unaffected by poor weather, extreme weather
conditions such as thunderstorms or very high winds can produce data
which can not be properly calibrated given the observing strategy that
we had used.  Based on our evaluation of the success of the
calibration procedure, we chose to delete a small number of epochs
from the light curves.  Paper I describes the reasons for flagging two
bad epochs in season 1.  In seasons 2 and 3 we flagged a total of 4
epochs due to bad weather conditions.  We also flagged 3 points
obtained at the very beginning of the seasons during which the VLA was
in a mixed configuration while the antennas were being moved from
their D-array positions to their A-array positions.  For these epochs,
the absolute flux density calibration was not correct.  The number of
epochs remaining after flagging the data from each season is listed in
the $N_{\rm good}$ column of Table~\ref{tab_obs}.

\subsection{Correction of Systematic Errors and Final Light Curves
\label{sec_finalcurves}}

Paper I and FT01 describe a method that uses the light
curves of steep-spectrum radio sources to correct for systematic
errors in the absolute flux density calibration.  We have followed the
same method in this paper to correct the light curves of the lensed
images.  The results presented in FT01 showed that CSOs are
excellent flux density calibrators at the frequencies used for our
monitoring observations.  As discussed in FT01, the light
curves of J1823+7938 and J1945+7055 showed evidence of additional
systematic effects not seen in the light curves of the other CSOs.
Thus, those two sources were not included in the construction of the
season 3 correction curve.  Additionally, the season 2 data show that
1633+741, although relatively stable, has noisier light curves than
those of the CSOs.  This is probably due to the significant extended
emission and complicated morphology of 1633+741.  Our snapshot
observations did not cover the $(u,v)$ plane well enough to accurately
model the complex emission of the source.  Thus, we dropped 1633+741
as a secondary flux density calibrator.

There were two details to consider in constructing the correction
curves.  The first was that the primary flux calibrator, 3C\,343 has
some extended emission.  The B-configuration observations are more
sensitive to lower surface brightness emission than are the
A-configuration observations.  Thus, the total flux density observed
for 3C\,343 will increase whenever the VLA enters the B configuration.
The observations of the secondary flux calibrators were used to
examine the size of this effect.  The effect only appeared to be
significant during season 3, perhaps because the brightness of the
central component had decreased slightly and the extended emission
thus contributed a larger percentage to the total B-configuration flux
density.  To ensure that the correction curves did not include any
step function introduced by the changing flux density of 3C\,343, we
computed for each season the mean flux densities of the CSOs in the A,
BnA, and B configurations separately and used those values to
normalize the light curves.  This procedure also corrected for any
step functions caused by the presence of low-level extended emission
in the secondary flux calibrators themselves.  The second detail to
consider was that, by dropping 1633+741, we had no secondary flux
density calibrators for the recalibrated season 1 data.  However, our
analysis of the light curves and the time delay results showed that
the inclusion of the 1633+741 data introduced more scatter than it
took out.  Thus, we believe that our decision to drop the 1633+741
data was the proper one.

\begin{figure}
\plotone{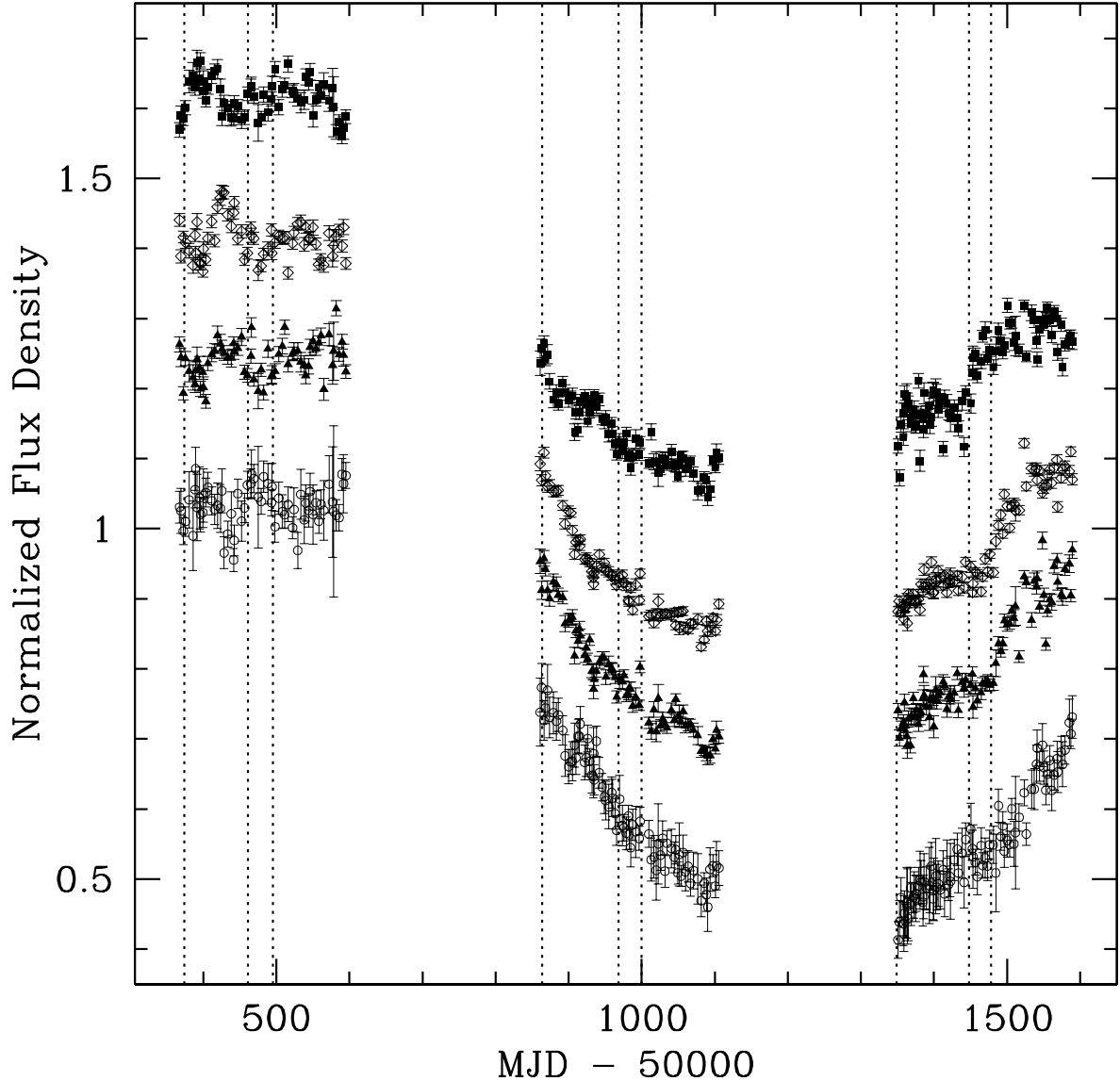}
\caption{Normalized light curves of the four lensed images of the
background radio source in the CLASS B1608+656 lens system obtained
over the three observing seasons.  From top to bottom, the curves
correspond to components B (filled squares), A (open diamonds), C
(filled triangles), and D (open circles).  The abscissa is time, with
units of MJD $-$ 50000.  Each curve has been normalized by its mean
flux density over the course of the observations, so the plotted
points represent fractional changes in the component flux densities.
For clarity, each light curve has been shifted vertically by an
arbitrary amount.  The component B light curve has been shifted by
$+$0.3, the component A curve by $+$0.08, the component C curve by
$-$0.08, and the component D curve by $-$0.30.  For each season, the
vertical dotted lines delineate the times at which the VLA
configuration changed: D$\rightarrow$A, A$\rightarrow$BnA, and
BnA$\rightarrow$B, from left to right.  The gaps in the curves
correspond to times in which the VLA was in the BnC, C, CnD, and
D-arrays, during which the angular resolution provided by the
array was not high enough to cleanly resolve the four components
in the system.
\label{fig_lc4comb}}
\end{figure}

The final corrected light curves for the four components in B1608+656
are shown in Figure~\ref{fig_lc4comb}.  The errors on the points are a
combination in quadrature of (1) the RMS noise in the residual map
from the {\tt DIFMAP} processing, typically $\sim$0.1 mJy~beam$^{-1}$,
and (2) a 0.6\% fractional uncertainty arising from errors in the
absolute flux density calibration.  We have estimated the fractional
uncertainty from the season 3 CSO light curves.
All of the light curves have been normalized by their mean flux
densities over the entire three seasons of observations. Thus, the
plot in Figure~\ref{fig_lc4comb} represents the fractional variations
in flux density, which allows for direct comparison of the component
light curves.  The figures show that the background source varied
significantly in both season 2 and season 3.  The variations in each
of these seasons are on the order of 25--30\%, in contrast to the
$\sim$5\% variations seen in season 1.  Unfortunately, the season 2
light curves have nearly a constant slope, which made it difficult to
determine unambiguously the time delays and magnifications needed to
align the component light curves.  However, the season 3 light curves
have both a large amount of variation and clear changes in slope,
making them excellent inputs for determining time delays.

\section{Determination of Time Delays}

\subsection{Dispersion Method}

In Paper I we used several statistical methods to determine the delays
between the light curves.  In this paper, with obvious source
variability and $\sim$250 points in the light curves of each
component, we have used only the dispersion method described by
\citet{pelt1,pelt2}.  This method has the clear advantage of not
requiring any interpolation of the input light curves.  As we did in
Paper I, we used the $D^2_2$ and $D^2_{4,2}$ methods \citep[as defined
in][]{pelt2} to compare the three independent pairs of light curves.
In these methods, one of the input light curves is shifted in time by
a delay ($\tau$), scaled in flux by a relative magnification ($\mu$),
and combined with the second curve to create a composite light curve.
The internal dispersion in the composite curve is then calculated by
computing the weighted sum of the squared flux difference between
pairs of points in the curve.  For the $D^2_2$ method, only adjacent
pairs of points are considered, while, for the $D^2_{4,2}$ method,
additional pairs of points are considered.  In the $D^2_{4,2}$
calculations, all pairs separated by fewer than $\delta$ days
contribute to the final dispersion, but with an additional weighting
such that the pairs with the smallest separation in time have the
highest weights.  By including more pairs, the $D^2_{4,2}$ statistic
is less likely to be affected by individual discrepant points.

We searched a range of delays and relative 
magnifications to determine the values that, when applied to one of
the input curves, produced the best alignment between the two curves
(i.e., the lowest internal dispersion in the composite curve).
The $(\mu,\tau)$ pairs used for the search formed regular grids.  The
$\tau$ values were separated by 0.5~d in the range $\tau_0 \pm 30~{\rm
d}$.  The values of $\tau_0$ were those derived in Paper I, namely 31,
36, and 76~d.  The range of relative magnifications searched was
$0.95\mu_0$ to $1.05\mu_0$, with steps of size $\Delta \mu =
0.001\mu_0$.  The values of $\mu_0$ were also chosen to match the
best-fit values obtained from the season 1 data: 2.042, 1.038, and
0.351 for $\mu_{AB}$, $\mu_{CB}$, and $\mu_{DB}$, respectively.  Here,
we have defined $\mu_{AB}$ as the relative magnification of component
A with respect to component B, and so forth.  These relative
magnifications express the ratios of the component flux densities, once
the light curves have been properly corrected for the time delays.

\subsection{Delays from Combined Analysis \label{sec_comb_delay}}

As a first step in the determination of the delays, we analyzed the
results from each season individually.  The results from season 2 were
not conclusive due to the degeneracy between the relative magnification
and the time delay.  However, the season 1 and season 3 light
curves gave robust determinations of the time delays and relative
magnifications between the curves.  The delays determined from the two
seasons were consistent, but the magnifications were discrepant.  It
is possible that the absolute flux density scale might change from
season to season.  However, such a change would manifest itself as a
constant scaling applied to each of the component flux densities
equally, leaving the relative magnifications unaffected. We thus
concluded that some slow process had affected the measured flux
density of at least one of the images (see \S\ref{microlens}).  As a
result, our approach to the combined analysis of the data from all
three seasons changed.  Instead of finding a set of best-fit global
magnifications, we let the relative magnifications vary from
season to season.  Because the determination of the dispersion is an
additive process, the implementation of this change was
straightforward.  Consider the comparison of two light curves, $Q_i$
and $R_j$, which are sampled only during the three seasons defined by
the ranges $[t_{1,0},\ t_{1,f}]$, $[t_{2,0},\ t_{2,f}]$, and
$[t_{3,0},\ t_{3,f}]$.  The composite curve $T_k(t_k)$ was constructed
from the input curves, with the second curve shifted in time by a
delay $\tau$ and scaled by season-dependent relative magnifications $\mu_n$,
such that
\begin{equation}
T_k = \cases{
 Q_i,         &if $t_k = t_i$ \cr
 R_j / \mu_1, &if $t_k = (t_j - \tau)$ and $t_{1,0} \leq t_j \leq t_{1,f}$ \cr
 R_j / \mu_2, &if $t_k = (t_j - \tau)$ and $t_{2,0} \leq t_j \leq t_{2,f}$ \cr
 R_j / \mu_3, &if $t_k = (t_j - \tau)$ and $t_{3,0} \leq t_j \leq t_{3,f}$.
}
\end{equation}
The internal dispersion in the composite curve was then calculated in
the standard fashion described above.

\begin{figure}
\plotone{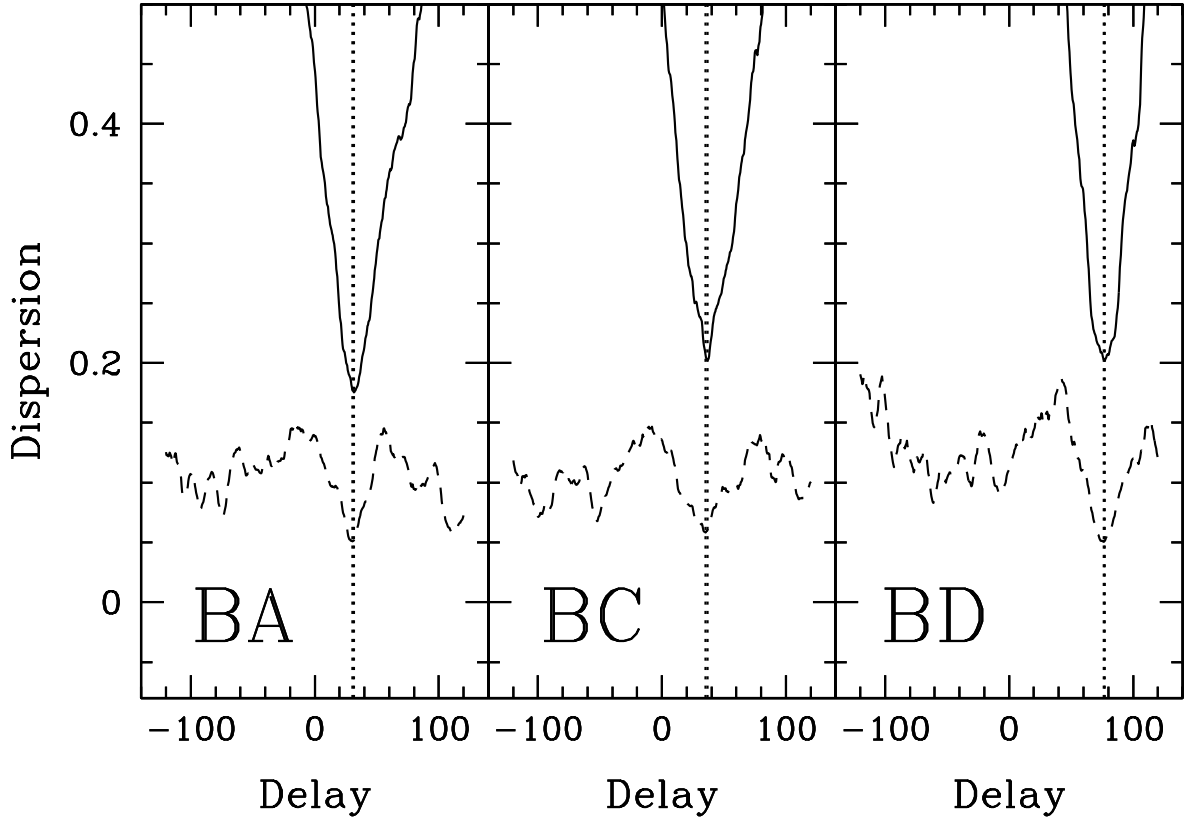}
\caption{Dispersion spectra from comparison of the light curves of
components B and A (left panel), B and C (middle panel), and B and D
(right panel).  The spectra were calculated with the $D^2_4$ method
with $\delta$ = 5.5 days (see \S\ref{sec_comb_delay}).  The curve
minima, marked with vertical dotted lines, indicate the time delays.
The solid curves represent the results from analysis of the combined
data sets from all three seasons, while the dashed curves represent
the results from season 1 alone.
\label{fig_pdisp3}}
\end{figure}

For each pair of curves, we searched a grid of $(\mu_1, \mu_2, \mu_3,
\tau)$ values to find the combination which gave the lowest
dispersion.  In order to ascertain the robustness of the derived time
delays, we repeated the $D^2_{4,2}$ method with a range of values for
$\delta$ and compared the results (Table~\ref{tab_combdisp}).
Figure~\ref{fig_pdisp3} shows typical output curves from the
dispersion analysis.  We show for comparison the equivalent curves
from the Paper~I analysis as dashed lines on the plots.  Although both
sets of curves give nearly the same time delays, the curves associated
with the combined data sets show much clearer minima.

\begin{center}
\begin{deluxetable}{lrrrrrrrrrrrrr}
\rotate
\tablenum{3}
\tablewidth{0pt}
\scriptsize
\tablecaption{Results of Dispersion Analysis of Combined Data Set
 \label{tab_combdisp}}
\tablehead{
\colhead{Statistic}
 & \colhead{$\delta$\tablenotemark{a}}
 & \colhead{$\tau_{BA}$}
 & \colhead{$\mu_{AB,1}$}
 & \colhead{$\mu_{AB,2}$}
 & \colhead{$\mu_{AB,3}$}
 & \colhead{$\tau_{BC}$}
 & \colhead{$\mu_{CB,1}$}
 & \colhead{$\mu_{CB,2}$}
 & \colhead{$\mu_{CB,3}$}
 & \colhead{$\tau_{BD}$}
 & \colhead{$\mu_{DB,1}$}
 & \colhead{$\mu_{DB,2}$}
 & \colhead{$\mu_{DB,3}$}
}
\startdata
$D^2_2$     & \nodata & 32.5 & 2.042 & 1.986 & 2.016 & 36.5 & 1.038 & 1.031 & 1.026 & 78.5 & 0.351 & 0.344 & 0.343 \\
$D^2_{4,2}$ & 3.5     & 32.0 & 2.042 & 1.982 & 2.010 & 35.5 & 1.038 & 1.032 & 1.026 & 78.5 & 0.351 & 0.344 & 0.342 \\
$D^2_{4,2}$ & 4.5     & 32.0 & 2.042 & 1.984 & 2.010 & 36.5 & 1.039 & 1.030 & 1.026 & 77.5 & 0.351 & 0.345 & 0.343 \\
$D^2_{4,2}$ & 5.5     & 31.5 & 2.042 & 1.984 & 2.008 & 36.5 & 1.039 & 1.030 & 1.027 & 80.0 & 0.351 & 0.344 & 0.344 \\
$D^2_{4,2}$ & 6.5     & 31.5 & 2.042 & 1.984 & 2.008 & 35.5 & 1.038 & 1.031 & 1.027 & 80.0 & 0.351 & 0.344 & 0.344 \\
$D^2_{4,2}$ & 7.5     & 31.5 & 2.042 & 1.984 & 2.006 & 35.5 & 1.039 & 1.031 & 1.028 & 80.0 & 0.351 & 0.344 & 0.344 \\
$D^2_{4,2}$ & 8.5     & 31.5 & 2.042 & 1.984 & 2.008 & 36.5 & 1.039 & 1.029 & 1.028 & 78.0 & 0.351 & 0.345 & 0.343 \\
$D^2_{4,2}$ & 9.5     & 31.5 & 2.042 & 1.986 & 2.006 & 36.5 & 1.039 & 1.030 & 1.029 & 78.0 & 0.351 & 0.345 & 0.343 \\
$D^2_{4,2}$ & 10.5    & 31.5 & 2.042 & 1.986 & 2.006 & 35.5 & 1.039 & 1.031 & 1.028 & 77.0 & 0.351 & 0.345 & 0.342 \\
$D^2_{4,2}$ & 11.5    & 31.0 & 2.044 & 1.986 & 2.004 & 35.5 & 1.039 & 1.031 & 1.028 & 77.0 & 0.351 & 0.345 & 0.343 \\
$D^2_{4,2}$ & 12.5    & 31.5 & 2.044 & 1.986 & 2.006 & 36.0 & 1.039 & 1.031 & 1.028 & 77.5 & 0.351 & 0.345 & 0.343 \\
$D^2_{4,2}$ & 13.5    & 31.0 & 2.044 & 1.988 & 2.006 & 36.0 & 1.039 & 1.031 & 1.029 & 77.0 & 0.351 & 0.345 & 0.342 \\
$D^2_{4,2}$ & 14.5    & 31.5 & 2.042 & 1.988 & 2.004 & 35.5 & 1.039 & 1.031 & 1.028 & 77.0 & 0.351 & 0.345 & 0.342 \\
$D^2_{4,2}$ & 15.5    & 31.5 & 2.044 & 1.986 & 2.006 & 36.0 & 1.039 & 1.031 & 1.029 & 76.5 & 0.351 & 0.345 & 0.342 \\
$D^2_{4,2}$ & 16.5    & 31.5 & 2.044 & 1.988 & 2.006 & 36.0 & 1.039 & 1.031 & 1.029 & 77.0 & 0.351 & 0.345 & 0.342 \\
$D^2_{4,2}$ & 17.5    & 32.0 & 2.042 & 1.986 & 2.006 & 36.0 & 1.039 & 1.032 & 1.028 & 76.5 & 0.350 & 0.345 & 0.342 \\
$D^2_{4,2}$ & 18.5    & 32.0 & 2.042 & 1.986 & 2.006 & 36.0 & 1.039 & 1.032 & 1.029 & 76.5 & 0.351 & 0.345 & 0.342 \\
$D^2_{4,2}$ & 19.5    & 31.5 & 2.042 & 1.988 & 2.004 & 35.5 & 1.039 & 1.032 & 1.028 & 76.5 & 0.351 & 0.345 & 0.342 \\
$D^2_{4,2}$ & 20.5    & 32.0 & 2.042 & 1.988 & 2.006 & 36.5 & 1.039 & 1.031 & 1.028 & 76.5 & 0.351 & 0.345 & 0.342 \\
\enddata
\tablenotetext{a}{The value of $\delta$ for
the $D^2_{4,2}$ method (see \S\ref{sec_comb_delay}).}
\end{deluxetable}
\end{center}

We have taken the median values of the columns in
Table~\ref{tab_combdisp} to be the best-fit time delays and relative
magnifications as determined from the data.  These values are given in
Table~\ref{tab_results}.  The median time delays are $\tau_{BA} =
31.5$~d, $\tau_{BC} = 36.0$~d, and $\tau_{BD} = 77.0$~d.  The time
delay ratios are $(\tau_{BD} / \tau_{BA}) = 2.44 $ and
$(\tau_{BD} / \tau_{BC}) = 2.14$.  These agree well
with the values obtained in Paper I.

\begin{figure}
\plotone{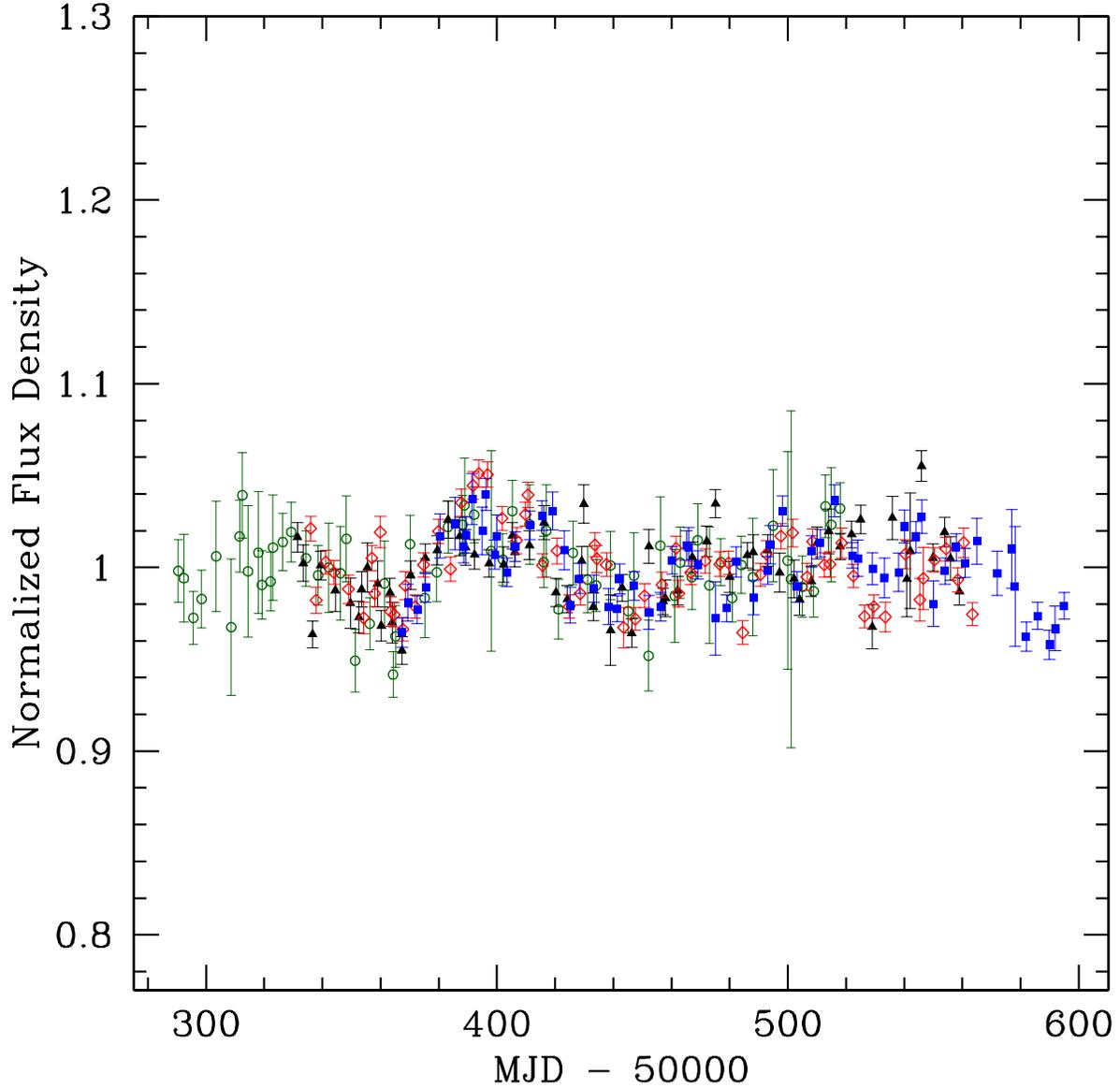}
\caption{Composite light curve constructed from the recalibrated data
from season 1.  The light curves for components A, C, and D have been
shifted by the time delays and relative magnifications given in
Table~\ref{tab_results} and overlaid on the component B light curve.
The resulting curve was then normalized by its mean value.  Component
D is the faintest image and thus the fractional measurement errors in
its flux density are larger than those of the other components.  The
data points associated with the four components are denoted by
open diamonds (red) for A, filled squares (blue) for B, filled
triangles (black) for C, and open circles (green) for D.  The vertical
scale is chosen to match those used in Figures~\ref{fig_compos_af340}
and \ref{fig_compos_ab922}.
\label{fig_compos_af310}}
\end{figure}

\begin{figure}
\plotone{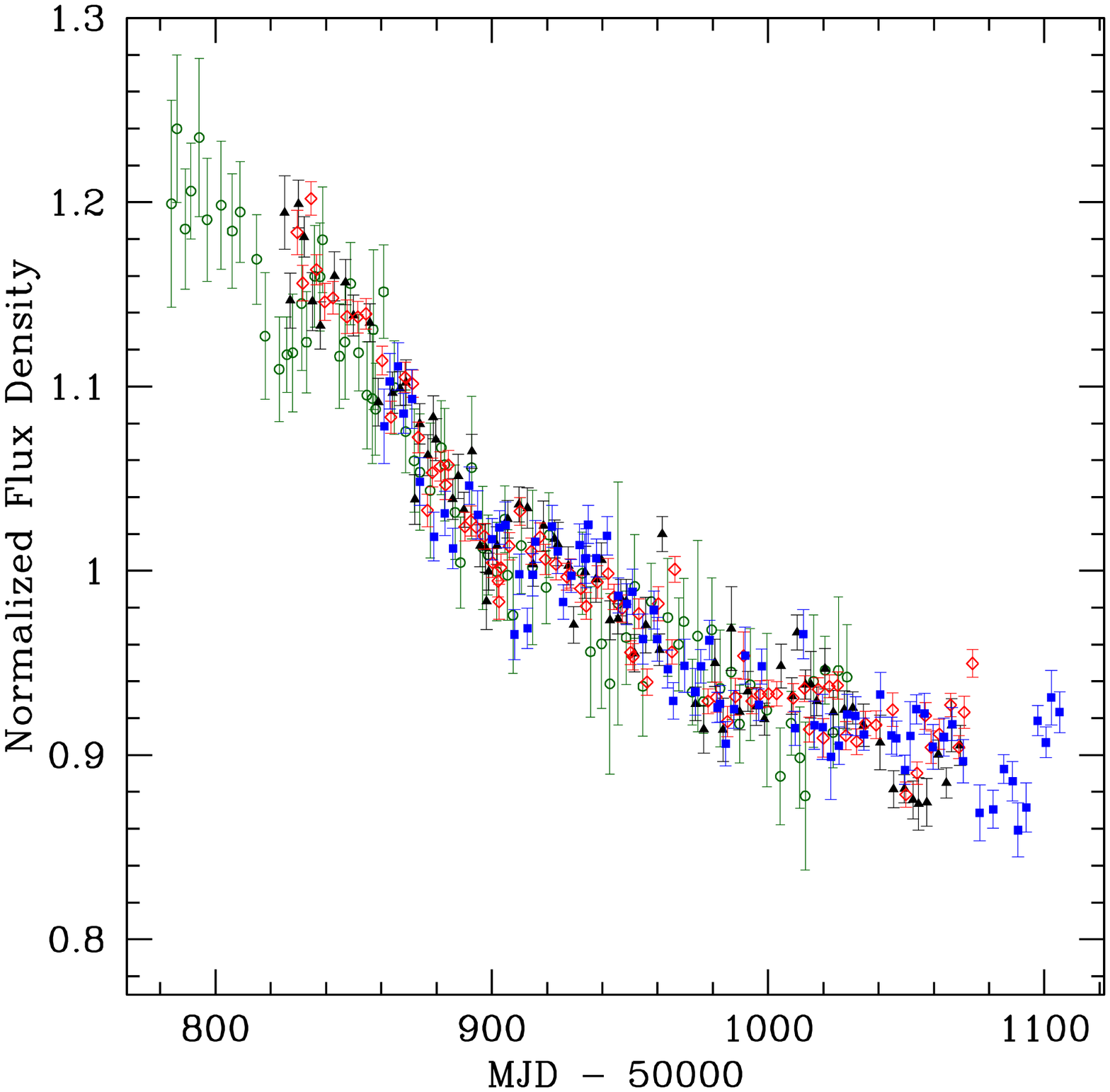}
\caption{Same as Figure~\ref{fig_compos_af310}, but for the data from
season 2.
\label{fig_compos_af340}}
\end{figure}

\begin{figure}
\plotone{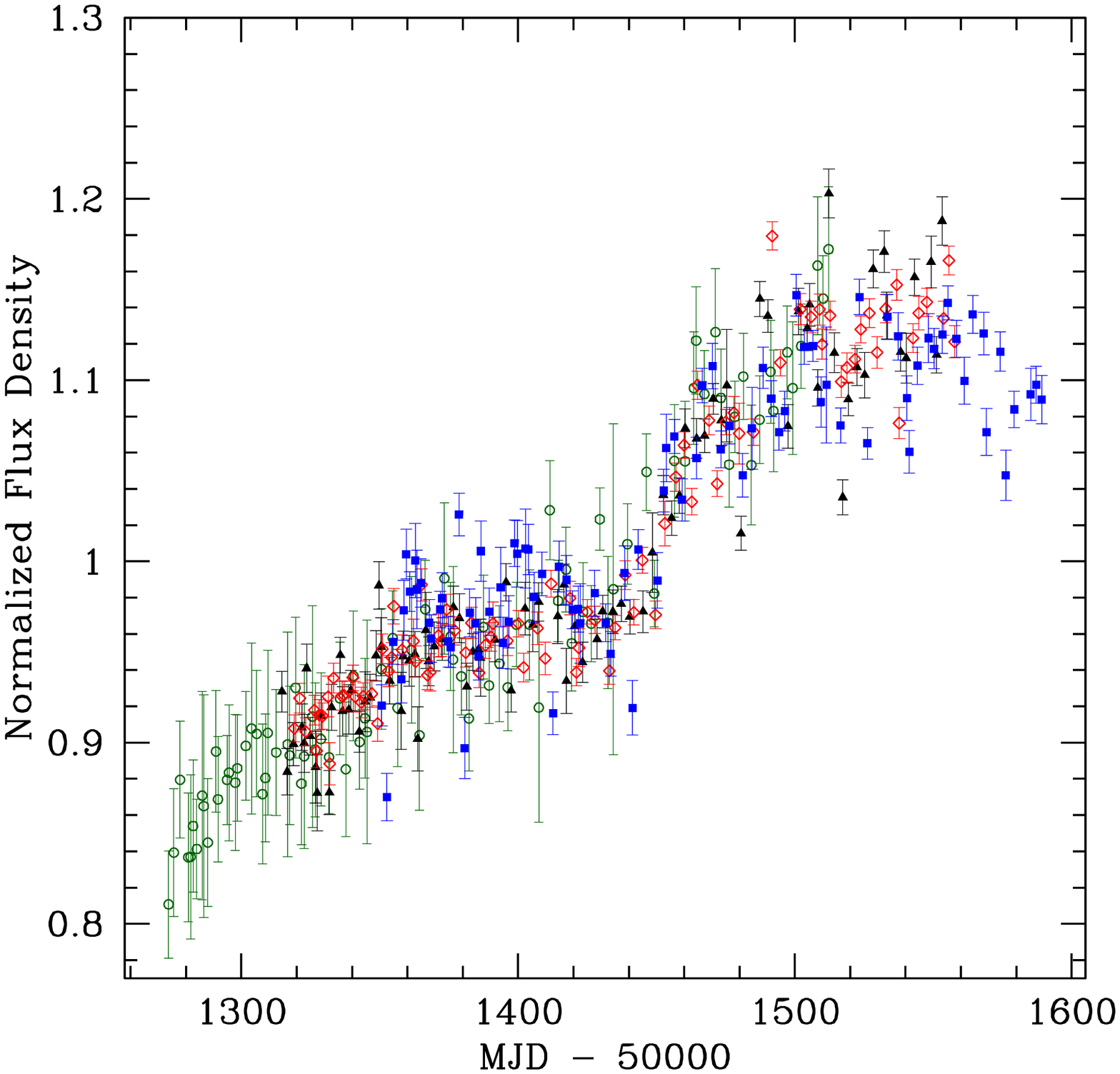}
\caption{Same as Figure~\ref{fig_compos_af310}, but for the data from
season 3.
\label{fig_compos_ab922}}
\end{figure}

We performed several consistency checks on the derived delays.  First,
we computed all possible delays between a leading component and a
trailing component, using the dispersion methods listed in
Table~\ref{tab_combdisp}.  The median values obtained for these delays
were $\tau_{AC} = 2.5$~d, $\tau_{AD} = 45.5$~d, and $\tau_{CD} =
44.0$~d.  These delays are consistent within the uncertainties
(\S\ref{sec_mc}) with the three delays calculated above.  Secondly, we
computed the delays between component B and the other three components
using the $\chi^2$-minimization technique described in Paper~I, which
requires that the input light curves be interpolated onto a
regularly-spaced grid.  The $\chi^2$-minimization results were
consistent with the dispersion method values, but had more
scatter about the median values.  The increased scatter is probably
due to small biases introduced by the various methods of interpolating
the input light curves.  The final check was to combine the input
light curves by shifting the A, C, and D curves by the appropriate
time delays and magnifications and then combining them with the
unshifted B light curve.  The resulting composite curves are shown in
Figures~\ref{fig_compos_af310}--\ref{fig_compos_ab922}.  The composite
curves show that the major features, and many of the minor features,
in each of the individual light curves are reproduced in the other
light curves at the proper times, giving additional confidence that
the measured delays are the correct ones.

\begin{center}
\begin{deluxetable}{cllll}
\tablenum{4}
\tablewidth{0pt}
\scriptsize
\tablecaption{Measured Quantities
\label{tab_results}}
\tablehead{
\colhead{Component}
 & \colhead{$\tau$\tablenotemark{a}}
 & \colhead{$\mu_1$\tablenotemark{b}}
 & \colhead{$\mu_2$\tablenotemark{b}}
 & \colhead{$\mu_3$\tablenotemark{b}}
}
\startdata
A & 31.5 & 2.042 & 1.986 & 2.006 \\
C & 36.0 & 1.039 & 1.031 & 1.028 \\
D & 77.0 & 0.351 & 0.345 & 0.342 \\
\enddata
\tablenotetext{a}{Time delay between component in table and
component B, in days.}
\tablenotetext{b}{Relative magnification of component in table with 
respect to component B, for season 1, 2, or 3.}
\end{deluxetable}
\end{center}

\section{Uncertainties on Measured Delays\label{sec_mc}}

We have estimated the uncertainties on the time delays and relative
magnifications through Monte Carlo simulations.  This is the same
method that was used in Paper I.  To summarize, we smoothed each of
the composite curves
(Figures~\ref{fig_compos_af310}--\ref{fig_compos_ab922}) to create the
input ``true'' light curves for the simulations.  One ``true'' light
curve was generated for each season, using the best-fit relative
magnifications appropriate for that season.
The delays used for each season were the global time delays derived
from the combined analysis (see Table~\ref{tab_results} for the input
delays and relative magnifications).  The smoothing method used to
generate the ``true'' curves from the composites was the
variable-width boxcar technique described in Paper I, with five points
in the smoothing window.  The simulated light curves were created by
adding a Gaussian-distributed random noise term to the points on the
``true'' curves.  The widths of the Gaussian distributions, $\sigma$,
were determined from the offsets between the measured composite curves
and their smoothed counterparts.  The values of $\sigma$ used ranged
from 1.4\% to 1.9\% of the mean flux densities of the lensed images.
We generated 1000 Monte Carlo realizations of the data associated with
each season.

\begin{figure}
%\figurenum{7a}
%\plotone{f7a.eps}
\plotfour{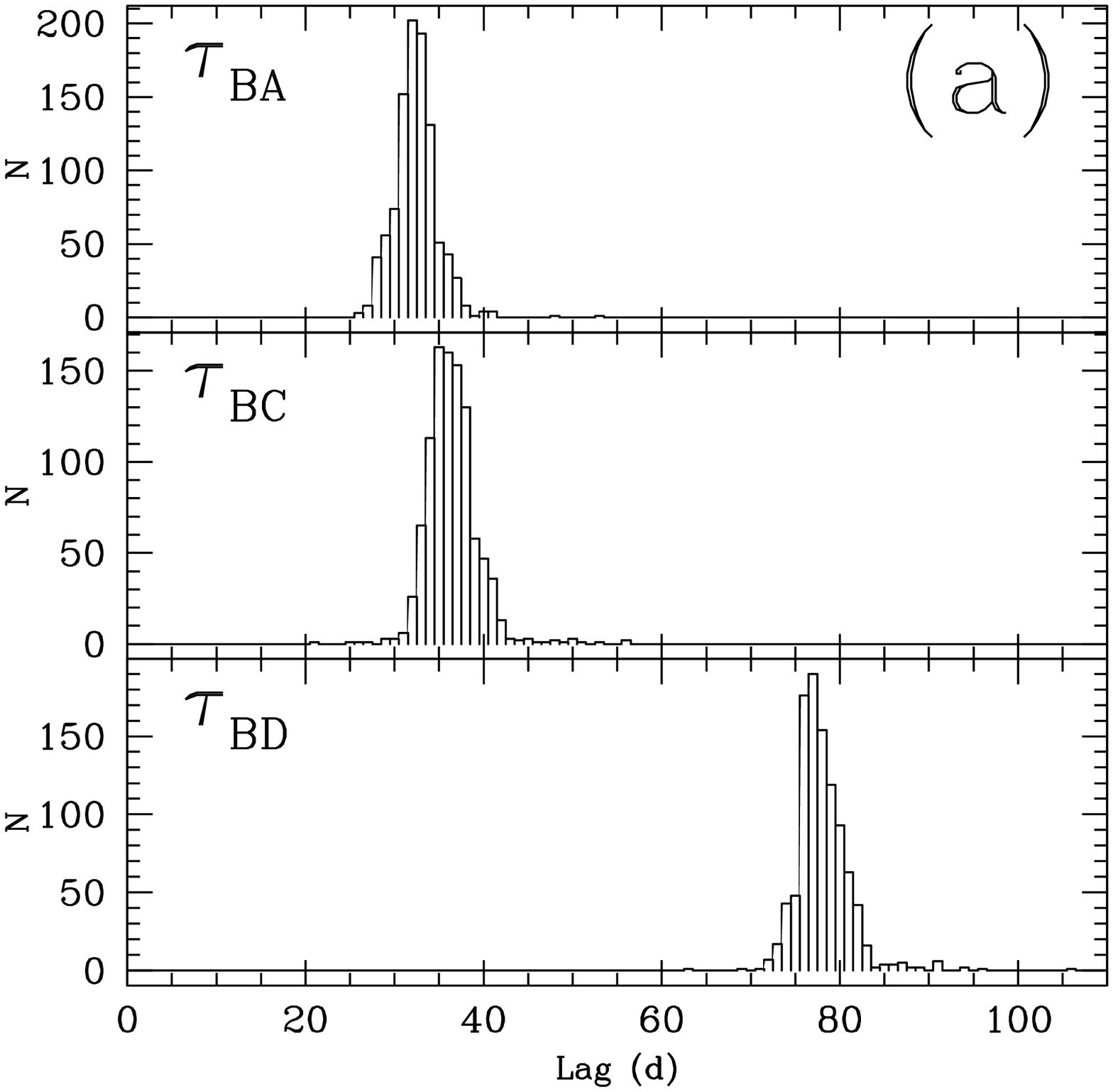}{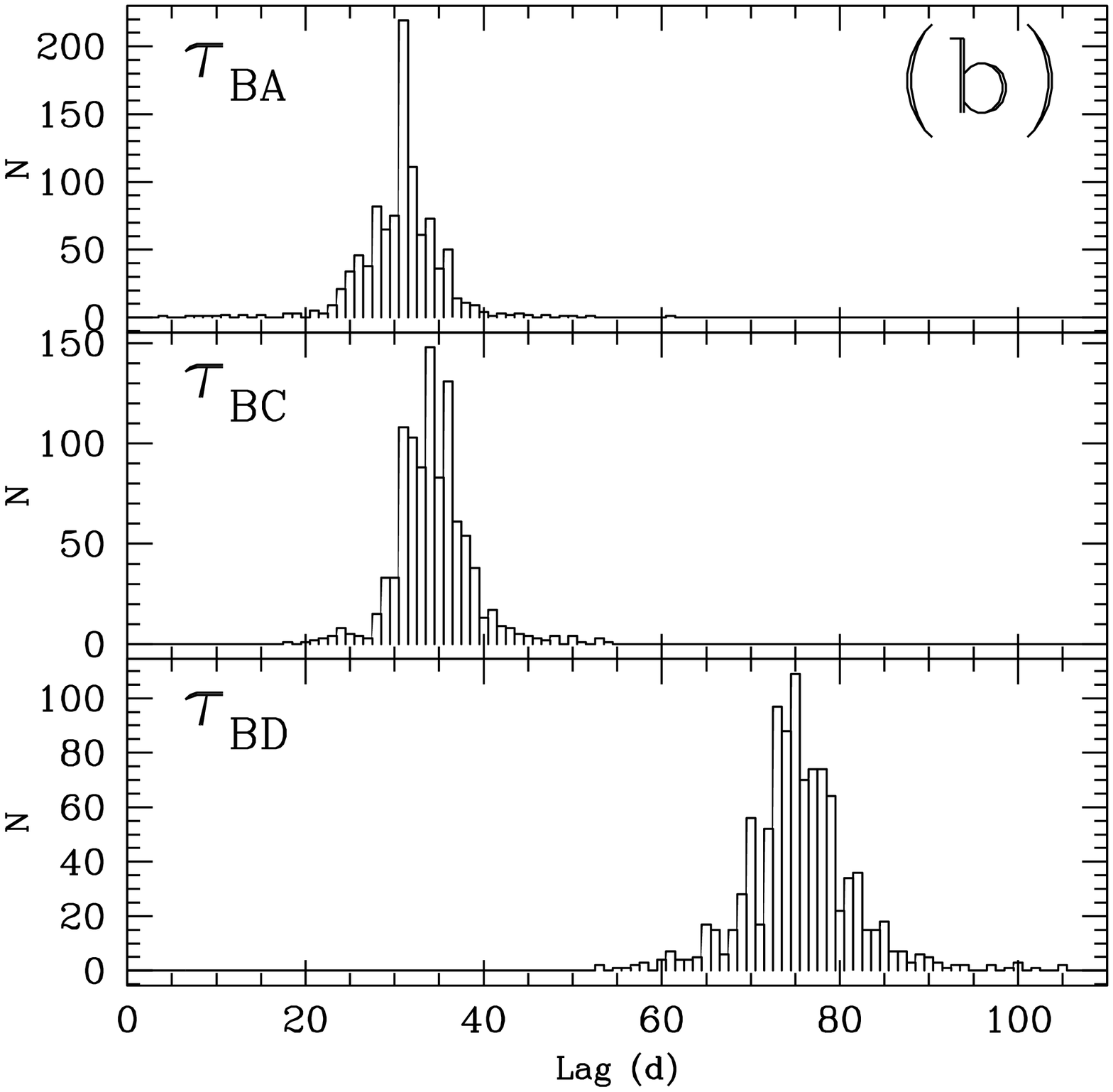}{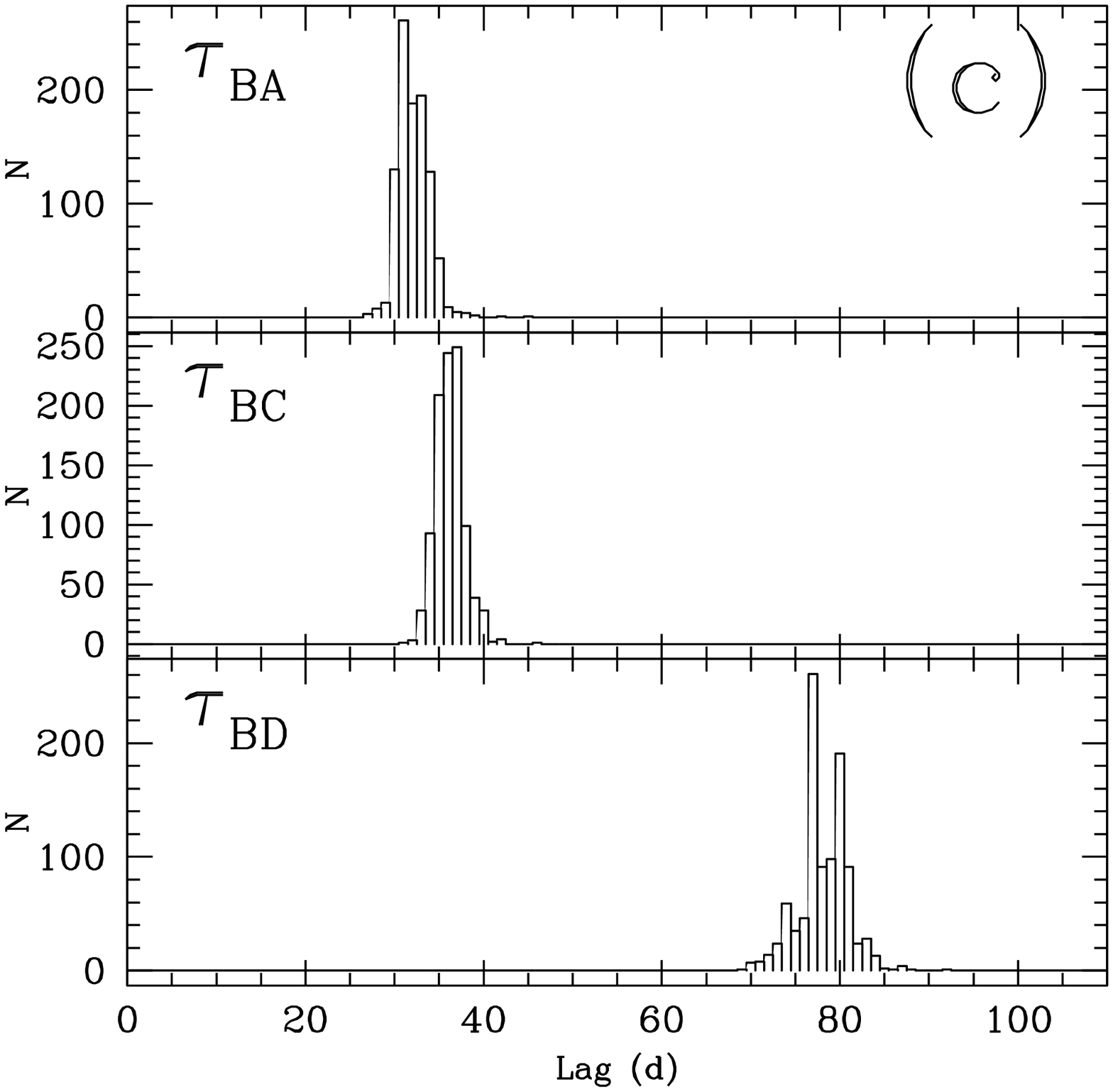}{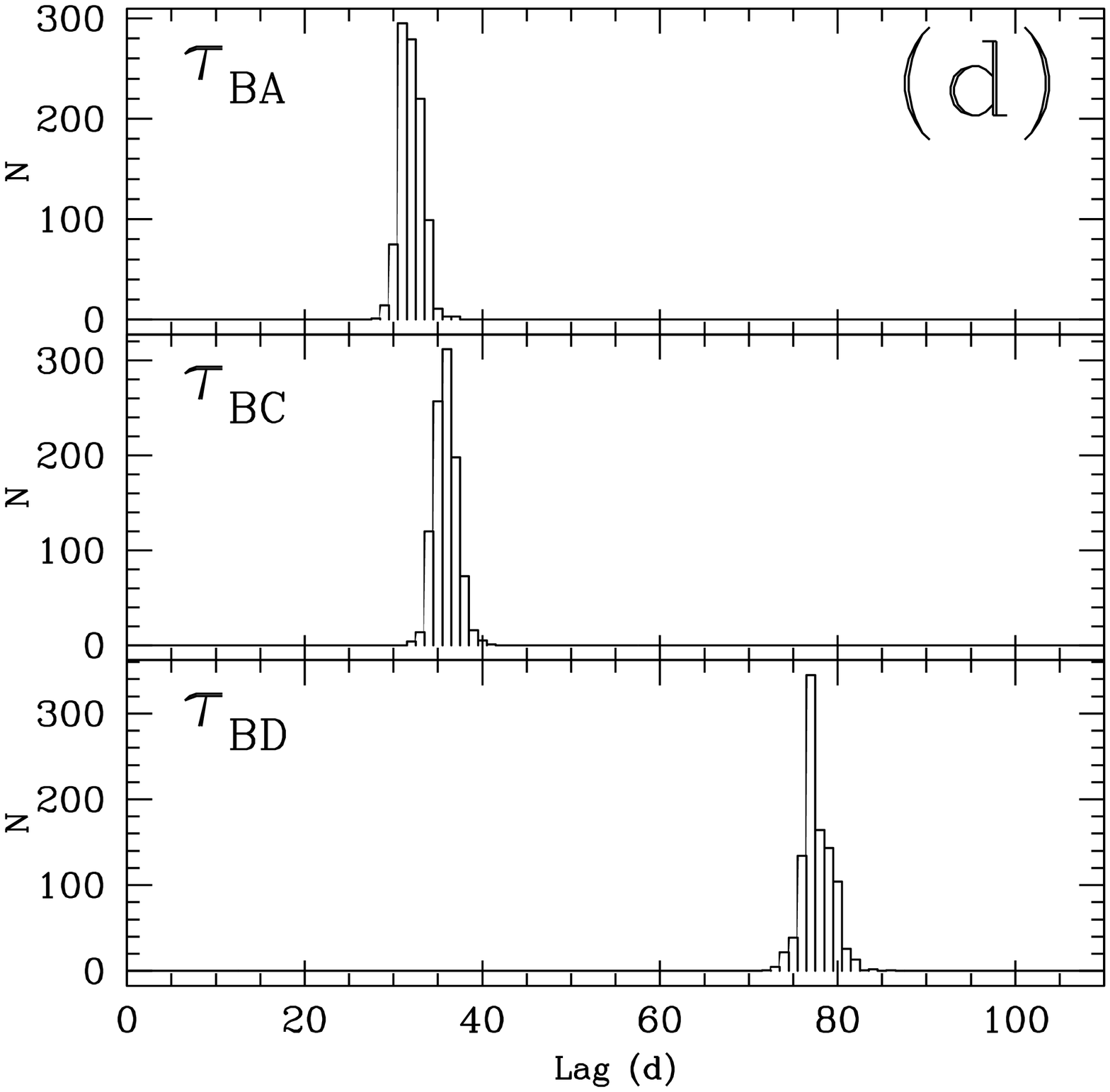}
\caption{Distribution of time delays resulting from dispersion-method
analysis of 1000 Monte Carlo realizations of the component light
curves.  (a) Season 1 results.  (b) Season 2 results.  (c) Season 3 results.
(d) Results from the analysis of the combined data from all three seasons.
\label{fig_mcdisp}}
\end{figure}

The simulated light curves were analyzed with the same code used for
the real light curves, using the $D^2_{4,2}$ method.  The analysis was
repeated using several different values of $\delta$, with no
significant difference in the results.  The results in
Table~\ref{tab_mcresults} were obtained for $\delta = 4.5$d.  One
advantage of the method used to analyze the monitoring data is that it
yielded not only the time delays derived from the entire combined data
set, but also the delays derived from each season individually.  Thus,
the effectiveness of the combined analysis can be evaluated.  The
distributions of time delays recovered from the simulations is shown
in Figure~\ref{fig_mcdisp}, where the first three figures show the
analyses of the individual seasons and the fourth shows the results
from the combined data sets.  The figures show that the distributions
are non-Gaussian and that the season 2 data set does not constrain the
time delays well.  The latter result was expected given the nature of
the light curves from that season.  However, the analysis of the
combined data set yields tight limits on the time delays, which are
approximately a factor of two improvement on the limits derived from
the season 1 data alone.  All of the delays are now determined to
$\pm$3~d or better (95\% CL).  The uncertainties on the delays and
magnifications, as well as those on derived quantities, are given in
Table~\ref{tab_mcresults}.

%\begin{figure}
%\figurenum{7b}
%\plotone{f7b.eps}
%\caption{
%\label{fig_mcdisp_af340}}
%\end{figure}
%
%\begin{figure}
%\figurenum{7c}
%\plotone{f7c.eps}
%\caption{
%\label{fig_mcdisp_ab922}}
%\end{figure}
%
%\begin{figure}
%\figurenum{7d}
%\plotone{f7d.eps}
%\caption{
%\label{fig_mcdisp_comb}}
%\end{figure}

\begin{center}
\begin{deluxetable}{lccc}
\tablenum{5}
\tablewidth{0pt}
\scriptsize
\tablecaption{Parameter Confidence Intervals from Combined Analysis
 \label{tab_mcresults}}
\tablehead{
\colhead{Quantity}
 & \colhead{Input Value}
 & \colhead{68\% Confidence Interval}
 & \colhead{95\% Confidence Interval}
}
\startdata
$\tau_{BA}$     & 31.5   & 30.5--33.5   & 29.0--34.5 \\
$\tau_{BC}$     & 36.0   & 34.5--37.5   & 33.5--38.5 \\
$\tau_{BD}$     & 77.0   & 76.0--79.0   & 74.0--81.0 \\
$\mu_{AB,1}$    & 2.042  & 2.038--2.050 & 2.032--2.054 \\
$\mu_{AB,2}$    & 1.986  & 1.976--1.990 & 1.968--1.996 \\
$\mu_{AB,3}$    & 2.006  & 2.000--2.014 & 1.992--2.020 \\
$\mu_{CB,1}$    & 1.039  & 1.035--1.041 & 1.032--1.044 \\
$\mu_{CB,2}$    & 1.031  & 1.027--1.034 & 1.023--1.037 \\
$\mu_{CB,3}$    & 1.028  & 1.025--1.032 & 1.021--1.035 \\
$\mu_{DB,1}$    & 0.351  & 0.350--0.352 & 0.349--0.353 \\
$\mu_{DB,2}$    & 0.345  & 0.344--0.347 & 0.342--0.348 \\
$\mu_{DB,3}$    & 0.342  & 0.341--0.344 & 0.340--0.345 \\
$(\tau_{BD}/\tau_{BA})$       & 2.44   & 2.37--2.47   & 2.34--2.51 \\
$(\tau_{BD}/\tau_{BC})$       & 2.14   & 2.12--2.19   & 2.09--2.22 \\
\enddata

\end{deluxetable}
\end{center}

\section{Discussion}

\subsection{Evidence for Microlensing and Lens Substructure\label{microlens}}

\begin{figure}
\figurenum{8}
\plotone{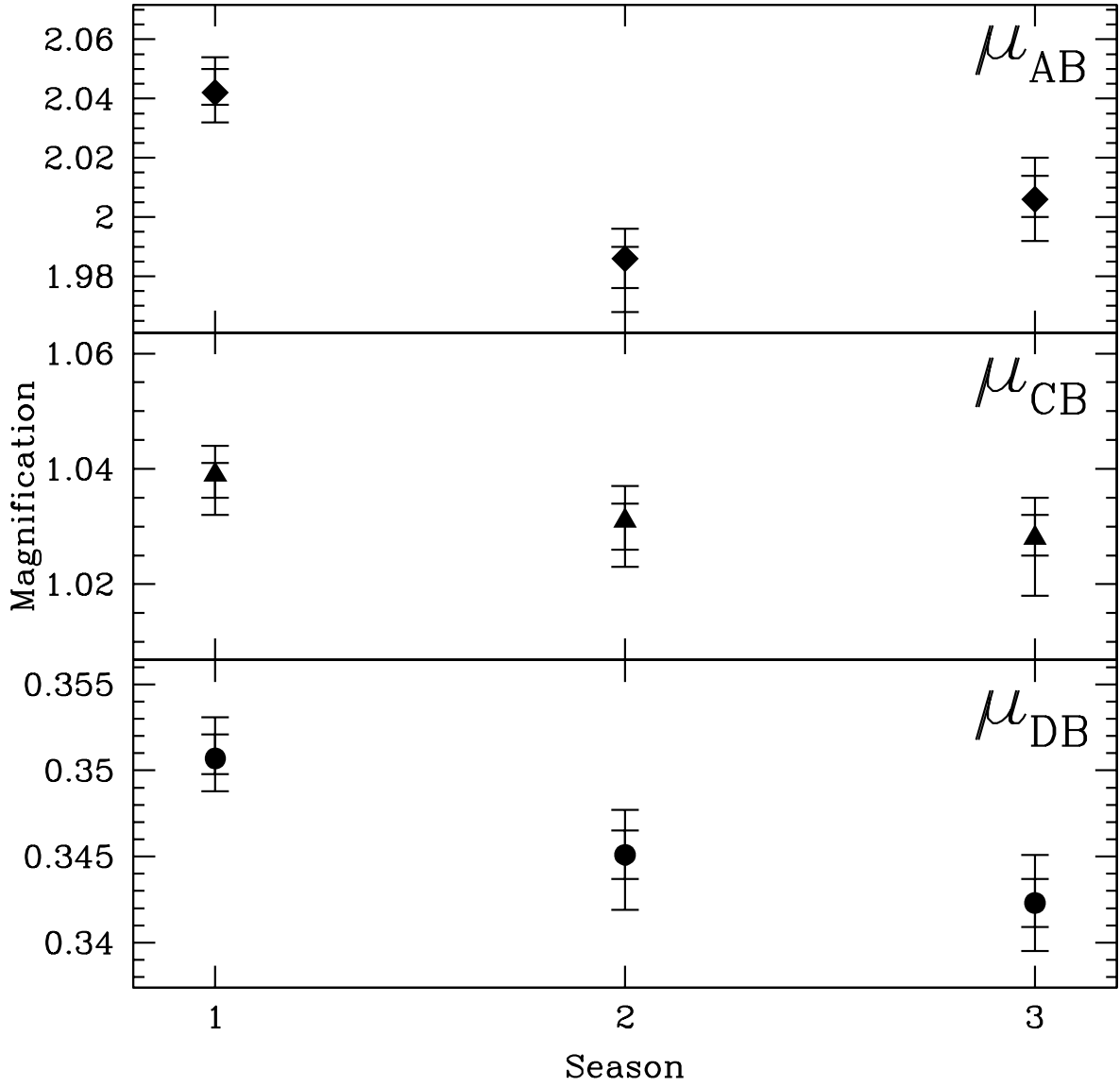}
\caption{Relative magnifications with respect to component B as a function
of the season of observations.  For each point, the inner error bars
are the 68\% confidence regions from the Monte Carlo simulations, while
the outer error bars are the 95\% confidence regions.
\label{fig_mulens}}
\end{figure}

As discussed in \S\ref{sec_comb_delay}, the component relative
magnifications were not constant from seasons 1 to 3.  In particular,
$\mu_{AB}$ and $\mu_{DB}$ decreased by 2\% or more during the course
of the observations (Figure~\ref{fig_mulens}).  Although not large,
these changes exceed the 95\% confidence limits on the relative
magnifications derived from the Monte Carlo simulations
(Table~\ref{tab_mcresults}), and thus appear to be real.  Similar or
more extreme changes in flux ratios are seen in optical monitoring of
lens systems \citep[e.g.,][]{2237micro,burud1600}.  These changes are
attributed to microlensing events, where stars or other massive
compact objects in the lensing galaxy change the magnifications of the
lensed images as they move through the galaxy.  For many years, it was
thought that radio observations of gravitational lens systems should
be unaffected by microlensing because the angular sizes of compact
radio cores, typically on the order of a milliarcsecond, are much
larger than the microarcsecond lensing cross sections of stars.
However, radio monitoring of the lens CLASS B1600+434 has revealed
changes in the component flux ratio that has been attributed to
microlensing \citep{1600micro1}.  In the case of B1600+434 the
unexpected microlensing was interpreted as being due to the
superluminal motion of a microarcsecond-sized component in the jet of
the background quasar across the complex caustic structure produced by
compact objects in the lens galaxy halo \citep{1600micro1}.  The
lensed source in the B1608+656 system is the core of a classical radio
double source \citep{snellen1608}.  Thus, it is certainly possible
that there are extremely compact jet components associated with
B1608+656 as well.  On the other hand, the changes in the flux density
ratios may have other explanations, such as scintillation.  However,
because we have monitored the system at only one frequency, we do not
have the clear discriminant between microlensing and scintillation
that multifrequency monitoring provides \citep{1600micro1}.  With
properly designed future observations, it may be possible to determine
the cause of the changing flux density ratios in this system.

We note that, whatever the cause of the variability in the relative
magnifications, it may also be necessary to invoke the presence of
substructure in the B1608+656 lensing galaxies to explain the observed
flux densities of the images.  Simple, or even fairly complex, lens
models of the B1608+656 system have not been able to properly
reproduce all of the observed flux density ratios \citep[Paper
II;][]{sm1608}.  As we noted in Paper II, this discrepant
magnification could be produced by perturbations to the smooth mass
distributions assumed for the lensing galaxies.  The presence of such
substructure has been invoked to explain similarly discrepant flux
ratios in other four-image lenses, although in those cases the
incongruous ratios were those between bright merging images.  In some
cases, the substructure has been interpreted as globular clusters or
plane density waves \citep{substruct}.  Recently, several papers have
suggested that the substructure is due to the small dark matter
satellite halos expected from CDM structure-formation models
\citep[e.g.,][]{cdmstruct1,cdmstruct2,cdmstruct3}.

\subsection{Determination of $H_0$ with B1608+656}

To determine $H_0$ from a lens system, the measured time delays are
combined with those predicted by the model.  The predicted delays
depend on the values of the fundamental cosmological parameters, so
one must assume a cosmological model.  Specifically, the model delays
are proportional to $(D_\ell D_s) / D_{\ell s}$ where $D_\ell$, $D_s$,
and $D_{\ell s}$ are the angular diameter distances from the observer
to the lens, from the observer to the source, and from the lens to the
source, respectively.  The angular diameter distances are functions of
$H_0$, $\Omega_m$, and $\Omega_\Lambda$, as well as the redshifts of
the lens and the background source.  We assume $(\Omega_M,
\Omega_\Lambda)$ = (0.3,0.7) for the rest of this paper.  In
comparison to this assumed cosmology, the value of $H_0$ derived from
this lens system will change by no more than $\sim$6\% in both flat
and open cosmological models with $0.1 < \Omega_M < 1.0$.  As specific
examples, the values of $H_0$ discussed below should be scaled by a
factor of 1.02 if $(\Omega_M,\Omega_\Lambda) = (0.3,0.0)$ and a factor
of 0.94 if $(\Omega_M,\Omega_\Lambda) = (1.0,0.0)$.

The considerable improvement in the accuracy of the measured time
delays between the three image pairs in B1608+656 warrants an update
of the lens model of this system, first presented in
Paper~II. Multiple time delays in a single lens system add additional
constraints on the lens model, and an improvement in their accuracy
could therefore lead to a change of the lens model and a change of the
inferred Hubble Constant.  Our models are based on the {\em lensmodel}
package developed by \citet{lensmodel}, but the results are consistent
with those based on the code used in Paper~II.  We emphasize that the
models presented here are only an update of the isothermal lens mass
models in Paper~II and that a much more detailed mass model will be
presented in a forthcoming publication, where we fully exploit
additional data on the system (see \S\ref{summary}).

For the lens modeling in this paper, we use the same constraints as in
Paper~II. We use the VLBI image positions with formal 1$\sigma$ errors
listed in Paper~II. Because the flux density ratios appear to slowly
change in time (\S\ref{sec_comb_delay}), even after the time-delay
correction of the light curves, we widen the error bars to 20\% on the
flux density ratios to err on the side of caution. We use the improved
time delays presented in this paper with 1.5~d errors (68\% CL). We
use singular isothermal ellipsoid (SIE) mass models for both lens
galaxies (G1 and G2; see Paper~II for details) and center them on the
galaxy centroids measured from {\em Hubble Space Telescope} (HST)
images obtained with the F555W, F814W, and F160W filters. The three
centroids differ by amounts that are significant, which could be due
to the presence of dust extinction and/or PSF problems.  In addition,
it is possible that the two merging lens galaxies are contained within
a common dark matter halo and, thus, that the luminous material may
not provide an accurate representation of the shape and centroid(s) of
the mass distribution.  These issues will be examined in more detail
in the forthcoming publication. Here, we present models for the
centroids from the F814W and F160W images; we reject the model based
on the F555W centroids because it produces an extra image of the
background source that is not seen in the data.  As already shown in
Paper~II, the differences in the centroids do not strongly affect the
inferred value of $H_0$, although the cause of the
wavelength-dependent shifts in position is something that requires
further study.

The mass distributions of G1 and G2 are allowed to have a free mass
scale (i.e., velocity dispersion), position angle, and ellipticity. At
this point no external shear is allowed.  Both lens galaxies are
assumed to be at the same redshift. The results of the updated lens
models and inferred values of $H_0$ are listed in
Table~\ref{tab_modparams}.  We note that these models are equivalent
to models I and II in Table~3 of Paper~II, in spite of using
completely independent modeling codes in the two papers.  We also note
that, in order to compare the models produced by the two codes, it is
important to understand how the mass scales in the models depend on
the projected axial ratio of the lensing galaxies, $q$.  We find that
the relation between the quantities representing the lens strengths,
$\sigma$ in Paper II and $b^\prime$ in Table~\ref{tab_modparams}, is
given by $b^\prime \sqrt{1+q^2} \propto \sqrt{q}\,\sigma^2$. Using
this relation, we find that the mass scales presented in Paper II and
in Table~6 are also equivalent and that the mass ratio between
galaxies G1 and G2 is $(M_1/M_2) \sim (b^\prime_1/b^\prime_2)^2
\sim3$. This large mass ratio avoids the creation of additional images
between galaxies G1 and G2.

We have also modeled the system with an allowance for an external
shear ($\gamma_{\rm ext}$).  The addition of shear to the models in
Table~\ref{tab_modparams} leads to a considerable decrease in the
value of $\chi^2$, for $\gamma_{\rm ext}\sim0.1$.  However, the number
of free parameters in the models increases to 11 (including $H_0$),
and the models are less well constrained than the models with no
shear.  Additionally, although there are several small galaxies in the
field around B1608+656, there is no evidence for a more massive group
or cluster that could yield a 10\% external shear, as for example in
the case of PG1115+080 \citep{kundic1115,tonry11151422}. Thus, we
think that the true external shear (i.e., a constant shear not due to
the lens galaxies themselves) is unlikely to be as high as 10\% and
that a revision of the galaxy mass models is more likely required.

The formal statistical errors on $H_0$ in the models are $\pm$1~\ksm\
and $\pm$2~\ksm\ for the 1$\sigma$ and 2$\sigma$ confidence limits,
respectively. In Paper~II we estimated that the range of reasonable
variations in the shapes of the mass profiles of the lens galaxies
contributed an additional uncertainty of 30\% to the determination of
$H_0$ from this system.  This estimate may be overly conservative,
given the small scatter in mass profile slopes seen in other lens
galaxies (\S\ref{sec_otherlens}).  However, given the complex nature
of the lens system and the uncertain positions of the lensing
galaxies, we will use this estimate of the systematic modeling
uncertainties in this paper.  In \S\ref{summary}, we discuss methods
by which the modeling uncertainties for this system may be reduced.

We note that an alternative and independent approach to lens modeling,
namely the non-parametric method developed by \citet{nonparam},
produces estimates of $H_0$ and its systematic uncertainties that are
similar to those determined in this paper.  In comparison to the more
frequently used analytic modeling approaches, the non-parametric
method has the advantage of making fewer assumptions about the form of
the mass distribution in the lensing galaxy or galaxies.  Thus, it is
possible to explore a larger region of the model space than is usually
done with single analytic models and, as a consequence, to obtain a
perhaps more realistic estimate of the uncertainties due to modeling.
On the other hand, given the additional freedom in model parameters,
the non-parametric approach may produce some descriptions of the
lensing galaxy that do not have a physical meaning, in spite of the
constraints introduced to produce realistic galaxy models.
Therefore, a straightforward interpretation of the results may be
difficult.
% the non-parametric approach may produce some
% descriptions of the lensing galaxy that do not have a physical
% meaning, so a straightforward interpretation of the results may be
% difficult.  
\citet{nonparam} applied their approach to the B1608+656
system, using the image positions and time delays as constraints.  The
time delays used by \citet{nonparam} were from an early analysis of
the Season 1 data and differ slightly from those found in this paper.
However, these small differences should not significantly affect the
resulting values of $H_0$.  \citet{nonparam} found that $\sim$90\% of
their model reconstructions produced values of $H_0$ between 50 and
100~\ksm.  They also applied their non-parametric method to PG1115+080
and the results were combined with the B1608+656 results to produce
$H_0 = 61 \pm 11\ (\pm 18)$ \ksm at 68\% (90\%) confidence, for
$(\Omega_M,\Omega_\Lambda) = (1.0,0.0)$.  For the cosmological model
adopted in this paper, their value of $H_0$ would change to 65 \ksm,
consistent with our results.

\begin{center}
\begin{deluxetable}{cccccccccr}
\tablenum{6}
\tablewidth{0pt}
\scriptsize
\tablecaption{Mass Model Parameters
 \label{tab_modparams}}
\tablehead{

 & \colhead{$x_c$\tablenotemark{a}}
 & \colhead{$y_c$\tablenotemark{a}}
 & \colhead{$b^\prime$}
 & 
 & \colhead{P.A.\tablenotemark{b}}
 & \colhead{$x_s$\tablenotemark{a}}
 & \colhead{$y_s$\tablenotemark{a}}
 & \colhead{$H_0$}
 &  \\
\colhead{Filter}
 & \colhead{(arcsec)}
 & \colhead{(arcsec)}
 & \colhead{(arcsec)}
 & \colhead{$q$}
 & \colhead{(deg)}
 & \colhead{(arcsec)}
 & \colhead{(arcsec)}
 & \colhead{(\ksm)}
 & \colhead{$\chi^2$}
}
\startdata
% F555W & $+$0.544 & $-$1.060 & 0.70 & 0.78 & $+$63.4 & $+$0.059 & $-$1.081 & 66 & 12  \\
%      & $-$0.337 & $-$0.976 & 0.39 & 0.45 & $+$51.2 &          &          &    &     \\ 
F814W & $+$0.521 & $-$1.062 & 0.68 & 0.88 & $+$71.8 & $+$0.060 & $-$1.090 & 65 & 43  \\ 
      & $-$0.293 & $-$0.965 & 0.39 & 0.39 & $+$53.1 &          &          &    &     \\ 
F160W & $+$0.446 & $-$1.063 & 0.69 & 0.91 & $-$58.0 & $+$0.058 & $-$1.103 & 61 & 196 \\ 
      & $-$0.276 & $-$0.937 & 0.36 & 0.31 & $+$53.8 &          &          &    &     \\ 
\enddata

\tablenotetext{a}{Positions are given in Cartesian rather than
astronomical coordinates.}
\tablenotetext{b}{Position angles are defined in the astronomical
convention, in degrees measured east of north.}

\tablecomments{Lens models for B1608+656 consisting of two SIE 
mass distributions with no external shear. Columns 2--3 
indicate the galaxy centroids measured in the HST images obtained
with the listed filter; columns 4--6 indicate the lens strength
(see alpha models in Keeton 2001), axial ratio, and position angle
of the SIE mass distribution; columns 7--8 indicate the source 
position; columns 9--10 indicate the inferred value of $H_0$ 
and the model $\chi^2$ value. The values of $H_0$ are quoted for 
$(\Omega_M,\Omega_\Lambda) = (0.3,0.7)$. Every first/second
line indicates the parameters for galaxy G1/G1.}
\end{deluxetable}
\end{center}

\subsection{Comparison to Determinations of $H_0$ with Other Lens Systems
 \label{sec_otherlens}
}

Time delays have been measured in 11 lens systems, with the B1608+656
delays being among those with the smallest uncertainties.
Unfortunately, the determination of $H_0$ from lenses is not as clean
as one might have hoped.  For nearly every lens with a measured time
delay, more than one model has been used to fit the data.  Often the
resulting values of $H_0$ are quoted with small formal errors that,
when taken at face value, make different determinations of $H_0$ with
the same lens system mutually exclusive at high levels of
significance.  These discrepant values of $H_0$ arise because
observations of most lens systems do not tightly constrain the mass
distribution of the lens, and there is a degeneracy between the
steepness of the lens mass profile and the derived value of $H_0$
\citep[e.g., Paper~II;][]{nonparam,witt_delays}.

Although it may not be possible to determine the radial mass profile
in many individual lens systems, there are indications that lensing
galaxies may follow a nearly universal mass profile.  It is, for
example, possible to assume that all lenses must give consistent
determinations of $H_0$ and then use the time-delay measurements to
obtain information about the mass distribution in the lenses.
\citet{cskdelays} has performed this experiment on a sample of five
lens systems with measured delays.  He finds that, in order to produce
the same value of $H_0$, the mass surface density properties of the
lensing galaxies must differ by only very small amounts.
Furthermore, several independent lines of investigation indicate that
not only are the mass distributions in many lenses similar, but that
the mass profiles are close to isothermal.  These approaches have used
models that incorporate more data than the standard two or four image
positions and fluxes.  The additional inputs make it possible to place
tight limits on the slope of the mass distribution.  The additional
constraints may come from full or partial Einstein rings
\citep[e.g.,][]{csk1654,keeton0957,cskrings}, complex structure in the
background source that in lensed into a ring-like configuration
\citep{cohn1933}, stellar dynamics \citep{tk2016,kt0047}, or the
orientations of mas-scale jets seen in very high angular resolution
radio maps \citep{rusin1152}.  Although it should not be concluded
that all lens galaxies have nearly isothermal mass distributions, it
may be that the range of mass profile slopes present in lensing
galaxies is significantly smaller than what is allowed by the
constraints in most individual lens systems with time delay
measurements.  If further observations strengthen these conclusions,
then the contribution of a major source of systematic error in
lens-derived measurements of $H_0$ may be significantly decreased.

The mass profile of the main lensing galaxy is not the only source of
uncertainty in the determination of $H_0$ from a lens system.  Another
factor enters if there is a cluster or massive group associated with
the primary lensing galaxy.  The inclusion of the effects of the
cluster can vastly complicate the lens model.  The case of the first
lens to be discovered, Q0957+561, is instructive.  Despite many years
of intensive observational and modeling efforts, no definitive model
has been developed.  For an excellent discussion of the wealth of
historical models of this system, see \citet{keeton0957}.  Clusters
have also been discovered in association with RX~J0911.4+0551
\citep{kneib0911,morgan0911} and possibly SBS~1520+530
\citep{faure1520}.  In addition, compact groups of galaxies have been
discovered in association with several lens systems
\citep[e.g.,][]{kundic1115,kundic1422,tonry11151422,tonry0751,tonry1131,fl0712,rusin1359},
although their effects on the lensing properties of the systems are
much smaller than those of clusters.  Still another problem in the
determination of $H_0$ from gravitational lens systems arises if the
location of the lensing galaxy is not known, as is the case in
B0218+357 \citep{lehar_2lenses}.

In spite of the possible problems mentioned above, the gravitational
lens method is attractive because many of the systematic errors
affecting the determination of $H_0$ with one lens system are
different from those affecting other systems.  Thus, it should be
possible to obtain a global measurement of $H_0$ with small
uncertainties by averaging the measurements from many lens systems.
This is in contrast to the distance-ladder techniques in which many of
the systematic uncertainties affect all distance determinations in the
same sense.  The major problem with the gravitational-lens method is
that the number of lens systems with measured time delays is still
small.  Even though the number of systems for which delays have been
measured has increased substantially over the last few years, some of
the delays are not measured to high precision.  In addition, lens
models for some systems are viewed with suspicion due to problems such
as those mentioned in the previous paragraph.  Thus, efforts to
determine a global value of $H_0$ often exclude a significant fraction
of systems with time-delay measurements.  Recent attempts have been
limited to samples consisting of five or fewer lenses, and have
obtained global values of $H_0$ ranging from $\sim$50 to $\sim$75~\ksm\
\citep[e.g., Paper~II;][]{schechreview,cskH0}.  With such small sample
sizes, the mean $H_0$ obtained can easily be biased by unknown factors
affecting one or two of the lens systems.  It is thus crucial to
measure time delays in more lens systems in order to obtain a robust
global determination of $H_0$ from lenses.

\subsection{Comparison to Determinations of $H_0$ with Other Methods}

There are, of course, methods for measuring $H_0$ that are completely
independent of the lens-derived values.  These include the traditional
distance-ladder methods in which Cepheid-based distances are used to
calibrate secondary distance indicators.  The HST Key Project combined
several secondary distance measurement methods to obtain a final value
of $H_0 = 72 \pm 3 (1 \sigma) \pm 7$~(sys) \ksm\ \citep{keyproject}.
Using an analysis of Cepheid-calibrated distances to Type Ia
supernovae, \citet{parodi_h0} derived $H_0 = 59 \pm 6$\ksm\ ($2
\sigma$ random errors combined with estimated systematic effects).
The use of the Sunyaev-Zeldovich (SZ) effect is another approach that,
like the gravitational lens method, is independent of the
distance-ladder approach.  In recent work by \citet{szh0}, the average
measurement obtained from a flux-limited sample of clusters was $H_0 =
66^{+14}_{-11} (1 \sigma) \pm 14$~(sys) \ksm.  The global values of
$H_0$ determined from lenses, the SZ effect, and traditional
distance-ladder methods are broadly consistent with each other and
with the $H_0$ determined from B1608+656.  The sources of systematic
error in each of these methods are different.  Therefore, if the
methods all produced values of $H_0$ that were in good, rather than
broad, agreement, the confidence that the correct value had been
measured would be increased.  It is important that steps be taken to
understand the systematic errors affecting each method and to attempt
to reduce those errors.  The systematic errors associated with the
gravitational lens method may be reduced if time delays can be
measured and mass profiles can be determined in significantly larger
samples of lens systems.

\section{Summary and Future Work
 \label{summary}
}

In this paper we have presented data obtained during three seasons of
monitoring of the B1608+656 gravitational lens system.  An analysis of
the combined data sets from the three seasons has led to an
improvement of factors of two to three in the precision of the three
time delays measured.  In addition, there is evidence that the
relative magnifications of the lensed images are changing.  Because
these relative magnifications should be independent of the absolute
flux density calibration, the observed changes may imply that at least
one of the images is being affected by microlensing.  If this is the
case, B1608+656 would be only the second lens system in which radio
microlensing has been detected.  A combination of the time delays with
revised lens models yields $H_0 =$ 61--65~\ksm, depending on the
positions used for the lensing galaxies, for
($\Omega_M,\Omega_\Lambda$) = (0.3,0.7). These values decrease by 6\%
for ($\Omega_M,\Omega_\Lambda$) = (1.0,0.0).  The uncertainties on
$H_0$ due to the time delay measurements are $\pm 1$ ($\pm 2$) \ksm\
for the 1$\sigma$ (2$\sigma$) confidence intervals.  Now that the time
delays have been determined to 4--10\% (95\% CL), the dominant source
of error in the determination of $H_0$ with the B1608+656 system comes
from the lens model.  The model uncertainties are on the order of $\pm
15$~\ksm.

There are two lensing galaxies within the ring of images in the
B1608+656 system, which presents a challenge to the modeling process.
Because of the disturbed nature of the system, discussions of general
time delay properties and the derivation of $H_0$ from lenses have
tended to discard B1608+656 from the samples being considered
\citep[e.g.,][]{witt_delays,cskH0}.  However, the B1608+656 system
does have the advantage of having a nearly complete Einstein ring that
is seen in optical and infrared HST
images \citep{disklenses,surpi1608}.  It thus may be possible to use
the ring data to constrain the mass profile slopes of the lensing
galaxies.  Techniques to incorporate the Einstein ring data into the
B1608+656 model have been described by \citet{rdb1608} and
\citet{cskrings}.  The observed ring is faintest between components B
and D, and is not detected at high significance in this region in the
available HST images \citep{cskrings,surpi1608}.  In order to fully
exploit the ring modeling techniques, it is critical to obtain deeper
HST images of the system so that the ring emission can be more clearly
separated from the noise.  The availability of the new Advanced Camera
for Surveys, with improved sensitivity and a finer pixel scale
compared to the Wide Field Planetary Camera 2, and the revived Near
Infrared Camera and Multi-Object Spectrometer will greatly aid in the
effort to get such images.  In addition to the analysis of Einstein
ring emission, information about the stellar dynamics of lensing
galaxies can place strong constraints on the mass distribution of the
lens \citep{tk2016,kt0047}.  We have obtained high resolution spectra
of the B1608+656 system with the Echelle Spectrograph and Imager on
the Keck~II telescope with which we will investigate the dynamics of
the lensing galaxies.  A full analysis of the Einstein ring structure,
the environment surrounding the lens system, and the stellar
kinematics should help to break the remaining degeneracies in the
model (e.g., precise galaxy positions, external shear, radial mass
profile, etc.) and, thus, significantly reduce the uncertainties on
$H_0$ due to the lens model.

%%%%%%%%%%%%%%%%%%%%%%%%%%%%%%%%%%%%%%%%%%%%%%%%%%%%%%%%%%%%%%%%%%%%%%%

\acknowledgments 
An observing program of this complexity could not have been completed
without the assistance of the NRAO staff.  Their efforts in keeping
the VLA running smoothly and their advice on the structure of the observing
program were invaluable.  We especially thank Meri Stanley, Jason
Wurnig, and Ken Hartley for making sure that all epochs had been
properly scheduled.  We are also grateful to the VLA operators for
implementing the observing program.  For useful discussions and
comments on the paper, we thank Ingunn Burud, Stefano Casertano, Tim
Pearson, Lori Lubin, Steve Myers, Frazer Owen, Rick Perley, Ken
Sowinski, and Greg Taylor.  We thank
the anonymous referee for suggestions that improved the paper.

%%%%%%%%%%%%%%%%%%%%%%%%%%%%%%%%%%%%%%%%%%%%%%%%%%%%%%%%%%%%%%%%%%%%%%%

\end{document}